\documentclass[aip,jmp,reprint,a4paper,amsmath,onecolumn]{revtex4-1}
\usepackage[utf8]{inputenc}
\usepackage[colorlinks=true,urlcolor=blue,anchorcolor=blue,citecolor=blue,filecolor=blue,linkcolor=blue,menucolor=blue,linktocpage=true,pdfa=true,unicode]{hyperref}
\usepackage{bm}
\usepackage{lmodern}
\usepackage{enumerate}
\usepackage{mathrsfs}
\usepackage{amssymb}
\usepackage{color}
\usepackage{ulem}
\newtheorem{theorem}{Theorem}
\newtheorem{definition}{Definition}

\newcommand{\DD}{\dfrac{\mathrm{d}^2}{\mathrm{d}x^2}}

\newcommand{\dom}[1]{\mathrm{Dom}\,#1}
\newcommand{\domd}[1]{\mathrm{Dom}\,#1^{\dagger}}
\newcommand{\Hil}{\mathcal{H}}
\newcommand{\inpr}[2]{\langle #1,#2\rangle}
\renewcommand{\emph}[1]{\textit{#1}}

\makeatletter
\newenvironment{equations}[1][]{\subequations\ifx\relax#1\relax\else\label{#1}\fi\align\ignorespaces}{
\endalign\ignorespacesafterend\endsubequations}
\def\@spliteq#1{\begin{equation}\begin{split}#1\end{split}\end{equation}}
\def\splitequation{\collect@body\@spliteq}

\makeatother

\begin{document}

\title{Self-Adjoint extensions of the one-dimensional Schrödinger operator with symmetric potential}

\author{Atsushi Higuchi}
\email{atsushi.higuchi@york.ac.uk}
\affiliation{Department of Mathematics, University of York, Heslington, York, YO10 5DD, United Kingdom}
\author{David Serrano Blanco}
\email{dsb523@york.ac.uk}
\affiliation{Department of Mathematics, University of York, Heslington, York, YO10 5DD, United Kingdom}

\date{\today} 

\begin{abstract}
We give an explicit correspondence between the domains of the self-adjoint extensions of a one-dimensional  Schrödinger differential operator with symmetric real-valued potential and the 
boundary conditions the functions in the resulting domains must satisfy. As is well known, each self-adjoint extension is parametrized by a unitary matrix.  We make the correspondence of this unitary matrix
with the boundary conditions explicit,  recovering the most familiar types of boundary conditions as special cases. 
We also demonstrate this correspondence implicitly for non-symmetric real-valued potential.
\end{abstract}


\maketitle

\section{Introduction}\label{Intro}

Formulation of Quantum Mechanics heavily relies on the theory of linear self-adjoint operators in a given Hilbert space. Indeed, quantum observables are defined to be self-adjoint operators acting on the space of states. Having a well defined set of self-adjoint operators is  essential in describing a mathematically rigorous and physically coherent quantum theory.  However, for most of the relevant systems in Quantum Mechanics, we do not start with a self-adjoint operator, but only with a symmetric one. The task is then to modify the given symmetric operator in such a way
that we end up with a self-adjoint operator which is related to the one we started with. 

From a mathematical point of view, an operator acting on a Hilbert space is defined via its action on the states and the domain on which it is allowed to act. One must make a clear distinction between an \emph{operation}, i.e., the action on states, and an \emph{operator}, 
i.e., the operation together with its domain. 
One must distinguish between a merely symmetric (or Hermitian) operator and a self-adjoint one having the same operation but defined on different domains. Construction of a self-adjoint operator from a symmetric operator consists in extending the original domain in a specific way. 

This process is known as constructing a \emph{self-adjoint extension} of an operator. It was originally introduced by Weyl~\cite{weyl} in the context of differential operators, and then generalized by von~Neumann~\cite{neu} for general linear operators defined on a Hilbert space. The main result, von~Neumann's theorem, states that every admissible self-adjoint extension is in one-to-one correspondence with the parameters of a unitary operation
on a certain space. The literature describing this procedure is extensive from a mathematical point of view~\citep{reed,kom,git}, and it has only recently been emphasized in physics related literature~\cite{ball,bonn,ara,ess,hall}. 

Even though the prescription given by von~Neumann is elegant and
self-contained, the explicit construction of self-adjoint extensions of symmetric operators is usually presented in the literature for specific examples. The explicit correspondence between 
the domain of the resulting self-adjoint operator and the set of boundary conditions associated with them is most commonly introduced as a convenient way to describe the domain instead of a consequence of 
the self-adjoint extension prescribed by von~Neumann's theorem.
One of the motivations for this paper is to explicitly derive the most general set of boundary conditions which uniquely characterize each self-adjoint extension resulting from von~Neumann's theorem for quantum particle in one dimension
with boundaries.

Thus, the aim of this paper is to give an explicit construction of the boundary conditions corresponding to the domains of the self-adjoint extensions given by von~Neumann's theorem of a one-dimensional Schrödinger operator that has a real-valued even potential. This example is sufficiently general to cover a broad set of physical systems. 
We have been led to analysing this system while applying von~Neumann's theorem
to the Laplace-Beltrami operator in  two-dimensional anti-de Sitter space for a scalar field theory.
As is well known, in anti-de Sitter space of any dimensions this Laplace-Beltrami operator
is not self-adjoint, but in certain cases it admits a family of self-adjoint extensions~\cite{ishi}.
The two-dimensional case for this theory has to be treated 
separately because there are two disjoint boundaries
unlike the higher-dimensional case and has not been studied in the context of von~Neumann's theorem so far in the
literature. This operator in two-dimensional anti-de Sitter space turns out to be  a one-dimensional Schrödinger-like operator with an even potential, and results of the present paper can be used to analyse it, as will be shown in a separate article.

The rest of the paper is organized as follows: In Sec.~\ref{Theory} 
we  review the general theory of self-adjoint extensions, starting with the relevant definitions, stating von~Neumann's theorem for operators in a general Hilbert space and concluding with an explicit prescription of 
self-adjoint extensions presented in a set of steps.
In Sec.~\ref{Body} we apply this prescription to a Schrödinger operator with a real-valued even potential and obtain its associated self-adjoint extensions described in terms of self-adjoint boundary conditions.
Finally in Sec.~\ref{BoundaryConditions} we classify the different possible boundary conditions corresponding to different choices of self-adjoint extension, which, by von~Neumann's theorem correspond to different choices of the unitary matrix associated with the self-adjoint extension.
We present details of some calculations in Appendices~\ref{appA} and \ref{appB}, and we give a brief outline to 
find the correspondence between the self-adjoint extensions given by von~Neumann's theorem and the
boundary conditions for the more general case of a non-symmetric potential in Appendix~\ref{appC}.

\section{The theory of self-adjoint extensions} \label{Theory}

We begin this section with elementary definitions and important results regarding (not necessarily bounded) linear operators on a Hilbert space. In this section we let $A$ be a linear operator defined on a Hilbert space $\mathcal{H}$ with inner product $\langle\cdot,\cdot\rangle$, and we will denote the domain of $A$ by $\dom{A}\subseteq\Hil$. 
\begin{definition}\label{d1}
The operator $A$ is said to be \emph{densely defined} if $\dom{A}$ is dense in $\Hil$, \it{i.e.}, that every element $f\in\Hil$ can be obtained as a limit of a sequence $\{f_n\}\in\dom{A}$. The operator $A$ is said to be \emph{closed} if, for any sequence $\{f_n\}\in\dom{A}$ satisfying $f_n\to f$ and $Af_n\to g$, it follows that $f\in\dom{A}$ and $Af=g$.
\end{definition}
Given a densely defined operator $A$ on $\Hil$, we define the subspace $\mathcal{D}$ as the space of all $\tilde{f}\in\Hil$ for which there exists an element $F\in\Hil$ such that
\begin{align}\label{1.1}
\inpr{\tilde{f}}{Af}=\inpr{F}{f}\,,
\end{align}
and that the linear functional $f\mapsto\inpr{\tilde{f}}{Af}$ is bounded~\cite{reed,hall} for every $f\in\mathrm{Dom}\,A$. 
\begin{definition}
Let $A$ be a closed, densely defined operator on $\Hil$, such that Eq.~\eqref{1.1} holds for every $f\in\dom{A}$. Then, the operator $A^{\dagger}$, with domain $\domd{A}:=\mathcal{D}$, defined by $A^{\dagger}\tilde{f}:=F$, is called the \emph{adjoint} of the operator $A$. Equivalently, the adjoint operator $A^{\dagger}$ of $A$ is uniquely characterized by
$\inpr{\tilde{f}}{Af}=\inpr{A^{\dagger}\tilde{f}}{f}$.
\end{definition}
Note that $\dom{A}$ being dense in $\Hil$ ensures that $\tilde{f}$ is uniquely determined by the relation \eqref{1.1}. 
\begin{definition}
A densely defined operator $A$ on a Hilbert space $\mathcal{H}$ is called \emph{symmetric} if 
$\dom{A}\subseteq\domd{A}$
and if $Af=A^{\dagger}f$  for all $f\in\dom{A}$. Equivalently, the operator $A$ is symmetric if and only if
\begin{align}\label{sy}
\inpr{f}{Ag}=\inpr{Af}{g}\,,
\end{align}
for all $f,g\in\dom{A}$. 
\end{definition}

It is important to note that the defining relation for a symmetric operator, namely, 
Eq.~\eqref{sy}, is only valid for elements in the domain of $A$. This fact is overlooked in some physics-oriented literature. Such an oversight may lead to apparent paradoxes in Quantum Mechanics (for concrete examples, 
see section 2 of Bonneau, et. al.~\cite{bonn} and Chapter I section 1.3 of Gitman, et. al.~\cite{git}). The following subclass of symmetric operators is extremely important in Quantum Mechanics.
\begin{definition}
The operator $A$ is called \emph{self-adjoint} if $A$ is symmetric and $\dom{A}=\domd{A}$.
\end{definition}
Now, suppose that $A$ is a self-adjoint operator on $\Hil$, and that there is an element $g\in\domd{A}=\dom{A}$ such that $A^{\dagger}g=\pm i g$. Then, since
$Ag=\pm i g$, it follows that
\begin{align}\label{1.2}
\mp i\inpr{g}{g}=\inpr{\pm i g}{g}=\inpr{Ag}{g}=\inpr{g}{A^{\dagger}g}=\inpr{g}{\pm i g}=\pm i\inpr{g}{g}\,,
\end{align}
and so $g=0$. This means that if $A$ is self-adjoint, then the equations $A^{\dagger}g=\pm i g$ cannot have non-trivial solutions $g\in \mathcal{H}$. This result is part of the proof of the following theorem.
\begin{theorem}{\bf : Basic criterion for self-adjointness.}
Let $A$ be a symmetric operator on $\Hil$. Then, the following statements are equivalent:
\begin{enumerate}[i)]
\item $A$ is self-adjoint.
\item $A$ is closed and $\mathrm{Ker}(A^{\dagger}\pm i\mathbb{I})=\{0\}$.
\item $\mathrm{Range}(A\pm i\mathbb{I})=\Hil$,
\end{enumerate}
where $\mathbb{I}$ stands for the identity operator in $\Hil$.
\end{theorem}
The implication (i)$\Longrightarrow$(ii) follows from the argument used in Eq.~\eqref{1.2}. For a detailed proof of the other implications see, for example, Chapter VIII of Reed and Simon~\cite{reed}. From this criterion it is clear that, for a given symmetric operator $A$, the spaces
\begin{align}
\mathscr{K}_{+}:=\mathrm{Ker}(A^{\dagger}-i\mathbb{I})\,,\hspace{.5cm}\mathscr{K}_{-}:=\mathrm{Ker}(A^{\dagger}+i\mathbb{I})\,, \label{K-plus-K-minus}
\end{align}
play an important role in determining whether or not $A$ is self-adjoint. These subspaces of $\Hil$ are called \emph{deficiency subspaces} associated to $A$, while the numbers $n_{\pm}:=\mathrm{dim}\,\mathscr{K}_{\pm}$ are called \emph{deficiency indices}.

The work we present in this paper concerns the \emph{self-adjoint extensions} of a symmetric operator.
\begin{definition}
Let $A$ and $A_U$ be linear operators on $\Hil$. The operator $A_U$ is said to be an \emph{extension} of $A$ if $\dom{A}\subset\dom{A_U}$, and $A_U f=Af$, for all $f\in\dom{A}$.
\end{definition}
An operator $A$ is said to be \emph{closable} if there exists a closed operator whose domain contains $\dom{A}$ and it has the same action of $A$, \emph{i.e.}, if $A$ admits a closed extension. The smallest closed extension of $A$, denoted by 
$\bar{A}$ is called the \emph{closure} of $A$. It is known that every symmetric operator is closable~\cite{hall}.
In this paper we study extensions of a symmetric operator $A$, 
denoted by $A_U$, which is self-adjoint. The criteria used to find such extensions is given by the next theorem originally proposed by Weyl ~\cite{weyl}, and then generalized 
by von~Neumann~\cite{neu}.
\begin{theorem}\label{t1}
Let $A$ be a closed symmetric operator on a Hilbert space, with deficiency indices $n_{+}$, $n_{-}$. Then,
\begin{enumerate}[i)]
\item the operator $A$ is self-adjoint if and only if $n_{+}=0=n_{-}$.
\item the operator $A$ has self-adjoint extensions if and only if $n_{+}=n_{-}$. The self-adjoint extensions of $A$ are in one-to-one correspondence with the unitary maps from $\mathscr{K}_{+}$ to $\mathscr{K}_{-}$.
\end{enumerate}
\end{theorem}
A proof can be found in standard literature ~\cite{reed,git}. 
In the next section we apply this theorem to discuss the self-adjoint extensions of the Schr\"odinger differential
operator in one dimension with symmetric real-valued potential on a finite interval. Before doing so, we need to elaborate on
how item (ii) of the above theorem works.  In this example there are two independent solutions to each of the equation
$A^\dagger g = \pm i g$.  That is, $n_{+}=n_{-}=2$.  A self-adjoint extension $A_U$ of $A$ is obtained by choosing
\begin{equations}
\dom{A_U}=\left\{f_0+g+Ug\,\left|\,f_0\in\dom{\bar{A}},\,g\in\mathscr{K}_{+}\right.\right\}\,,\label{1.5}\\
\hspace{.5cm}A_U(f_0+g+Ug)=\bar{A}^{\dagger}f_0+i g-i Ug\,,\label{1.6}
\end{equations} 
where $\bar{A}$ is the closure of the operator $A$ and where $U:\mathscr{K}_+\to \mathscr{K}_{-}$  
is a $2\times 2$ unitary matrix.  The operator $A_U$ thus defined is self-adjoint.

\section{Self-Adjoint extensions of the Schrödinger operator with symmetric potential}
\label{Body}
We 
consider the operator defined on $\Hil=L^2[-a,a]$, with $a$, a finite positive real number, by
\begin{align}\label{OpA}
A:=-\DD+V(x)\,,
\end{align}
where the real-valued piecewise-continuous potential satisfies $V(-x)=V(x)$, and such that $V(x)$ is a bounded operator on the Hilbert space $\Hil$~\cite{hall}. The domain of the operator is given by
\begin{align}
\mathrm{Dom}\,A=\left\{f\in AC^2[-a,a] \,\,|\,\, f(a)=f(-a)=f'(a)=f'(-a)=0\right\}\,,
\end{align}
where $AC^2[-a,a]$ denotes the set of functions  $f\in\Hil$ such that $f$ is differentiable, $f'$ is absolutely continuous~\cite{reed,royden}, and $f''\in L^{2}[-a,a]$. 
We choose the domain in this manner to ensure that $A$ is a closed, densely defined symmetric operator ~\cite{reed, kom}. It is clear that $A$ is not a self-adjoint operator because~\cite{reed,hall}
\begin{align}
\mathrm{Dom}\,A^{\dagger}=\left\{f\in AC^2[-a,a] \right\}\,,
\end{align}
that is, even though the formal expressions for $A$ and $A^{\dagger}$ as differential operators are identical, their domains are not the same.

We will now apply the prescription described in the previous section 
to find the self-adjoint extensions of $A$, and we start by finding its deficiency indices. 
By the assumption 
that  $V(x)$ is an even potential, we may choose the functions $g_{+},g_{-}\in AC^2[-a,a]$ 
as the normalized even and odd eigenfunctions of the equation
\begin{align}\label{defeq}
A\,g_{\pm}(x)=i g_{\pm}(x)\,.
\end{align}
Thus, $\mathscr{K}_{+}=\mathrm{span}\{g_{+},g_{-}\}$, 
where $\mathscr{K}_+$ is defined by Eq.~\eqref{K-plus-K-minus} and it immediately follows that $\mathscr{K}_{-}=\mathrm{span}\{\overline{g_{+}},\overline{g_{-}}\}$. Thus, we have $n_{+}=n_{-}=2$, and, hence, 
by theorem~\ref{t1}, the self-adjoint extensions of $A$ are 
parametrized by a $2\times 2$ unitary matrix.

Let $A_U$ denote the self-adjoint extension of $A$. Then, 
following \eqref{1.5}, we can explicitly write its domain as
\begin{align}\label{SAdom}
\mathrm{Dom}\,A_U=\left\{f_{\pm}\in AC^2[-a,a] \left|\,
\begin{matrix}
f_+(x)=g_+(x)+u_{11}\overline{g_+(x)}+u_{12}\overline{g_-(x)}\\[.5em]
f_-(x)=g_-(x)+u_{21}\overline{g_+(x)}+u_{22}\overline{g_-(x)}
\end{matrix}\,
\right.\right\}\,,  
\end{align}
where $u_{ij}$ are the $ij$-elements of a $2\times 2$ unitary matrix $U$.


Even though Eq.~\eqref{SAdom} specifies the domain of the self-adjoint extension $A_U$ of $A$ completely,
it is not in a form suitable for finding the spectrum of $A_U$.  It is more convenient to specify the self-adjoint extension
as a set of boundary conditions of functions in $AC^2[-a,a]$ at $\pm a$.  In this section we show 
how the specification of the domain
of $A_U$ given by Eq.~\eqref{SAdom} is translated into such boundary conditions.

Let us start by noting
that since the operator $A_U$ is symmetric, an element $\tilde{f}\in\domd{A_U} = \dom{A_U}$ satisfies 
\begin{align}
\inpr{\tilde{f}}{A_Uf}=\inpr{A_U\tilde{f}}{f}\,, \hspace{.3cm}\forall f\in\dom{A_U}\,,
\end{align}
where $\inpr{\cdot}{\cdot}$ is the standard $L^2[-a,a]$-inner product. Then  
by letting $f=f_\pm$, where $f_{\pm}$ is given by
Eq.~\eqref{SAdom}, we find
\begin{align}\label{2.1}
\inpr{\tilde{f}}{A_U(g_{j}+u_{j1}\overline{g_+}+u_{j2}\overline{g_-})}=\inpr{A_U\tilde{f}}{g_{j}+u_{j1}\overline{g_+}+u_{j2}\overline{g_-}}\,, \hspace{.5cm}j=1,2.
\end{align}
where we have let $g_{1}=g_{+}$ and $g_{2}=g_{-}$. Let us define
\begin{align}\label{G}
G_{j}(x):=g_{j}(x)+u_{j1}\overline{g_+(x)}+u_{j2}\overline{g_-(x)}\,,
\end{align}
so that, using integration by parts for the implicit integrals in the inner products we may write Eq.~\eqref{2.1} as
\begin{align}\label{G1}
\overline{G_{j}'(a)}\tilde{f}(a)-\overline{G_{j}(a)}\tilde{f}'(a)-\overline{G_{j}'(-a)}\tilde{f}(-a)+\overline{G_{j}(-a)}\tilde{f}'(-a)=0\,.
\end{align}
By adding and subtracting the cases $j=1$ and $j=2$ we arrive at the following pair of equations:
\begin{align}\label{2.2}
&\left[\overline{G_{1}'(a)}\pm\overline{G_{2}'(a)}\right]\tilde{f}(a)-\left[\overline{G_{1}'(-a)}\pm\overline{G_{2}'(-a)}\right]\tilde{f}(-a)\nonumber\\
&=\left[\overline{G_{1}(a)}\pm\overline{G_{2}(a)}\right]\tilde{f}'(a)-\left[\overline{G_{1}(-a)}\pm\overline{G_{2}(-a)}\right]\tilde{f}'(-a)\,.
\end{align}

Equation~\eqref{2.2} gives a pair of boundary conditions parametrized by a $2\times 2$ unitary 
matrix.  However, these
boundary conditions depend explicitly on $g_{\pm}(\pm a)$.   Next we show that this set of boundary conditions
is equivalent to another set of boundary conditions, which do no depend on the functions $g_{\pm}(x)$, 
parametrized also by a $2\times 2$ unitary matrix.
Let us define 
\begin{align}\label{fmatrices}
\vec{f}:=\begin{pmatrix}
\tilde{f}(a)+\tilde{f}(-a)\\[.5em]
\tilde{f}(a)-\tilde{f}(-a)
\end{pmatrix}\,,\hspace{.5cm}\vec{F}:=\begin{pmatrix}
\tilde{f}'(a)-\tilde{f}'(-a)\\[.5em]
\tilde{f}'(a)+\tilde{f}'(-a)
\end{pmatrix}\,.
\end{align}
Then, it can readily be verified, by using Eq.~\eqref{G} and remembering that $g_{+}$ ($g_{-}$) is an even (odd) 
function, that
\begin{align}\label{XYF}
X\vec{F}=Y\vec{f}\,, 
\end{align}
where
\begin{align}
X=\begin{pmatrix}
\overline{g_+(a)}+(\overline{u_{11}}+\overline{u_{21}})g_+(a) & \overline{g_-(a)}+(\overline{u_{12}}+\overline{u_{22}})g_-(a) \\[.5em]
\overline{g_+(a)}+(\overline{u_{11}}-\overline{u_{21}})g_+(a) & -\overline{g_-(a)}+(\overline{u_{12}}-\overline{u_{22}})g_-(a)
\end{pmatrix}\,,
\end{align}
and
\begin{align}
Y=\begin{pmatrix}
\overline{g_+'(a)}+(\overline{u_{11}}+\overline{u_{21}})g_+'(a) & \overline{g_-'(a)}+(\overline{u_{12}}+\overline{u_{22}})g_-'(a) \\[.5em]
\overline{g_+'(a)}+(\overline{u_{11}}-\overline{u_{21}})g_+'(a) & -\overline{g_-'(a)}+(\overline{u_{12}}-\overline{u_{22}})g_-'(a)
\end{pmatrix}\,.
\end{align}
Then, by defining
\begin{align}\label{ABM}
\mathcal{A}:=\begin{pmatrix}
g_+(a) & 0 \\[.5em]
0 & g_-(a)
\end{pmatrix}\,, \hspace{.5cm} \mathcal{B}:=\begin{pmatrix}
g_+'(a) & 0 \\[.5em]
0 & g_-'(a)
\end{pmatrix}\,,
\end{align}
and noting
\begin{align}
X=\begin{pmatrix}
1 & 1 \\[.5em]
1 & -1
\end{pmatrix}(\overline{\mathcal{A}}+\overline{U}\mathcal{A})\,,\hspace{.5cm}Y=\begin{pmatrix}
1 & 1 \\[.5em]
1 & -1
\end{pmatrix}(\overline{\mathcal{B}}+\overline{U}\mathcal{B})\,,
\end{align}
we can write Eq.~\eqref{XYF} as follows:
\begin{align}\label{3}
(\overline{\mathcal{A}}+\overline{U}\mathcal{A})\vec{F}&=(\overline{\mathcal{B}}+\overline{U}\mathcal{B})\vec{f}\,.
\end{align}
It is convenient to rearrange Eq.~\eqref{3} as
\begin{align}\label{5}
V(\vec{F}-i\vec{f}) = \tilde{V}(\vec{F}+i\vec{f})\,,
\end{align}
where
\begin{subequations}\label{Vs}
\begin{align}
V&:=\overline{\mathcal{A}}-i\overline{\mathcal{B}}+\overline{U}(\mathcal{A}-i\mathcal{B})\,,\label{V}\\
\tilde{V}&:=-\left[\overline{\mathcal{A}}+i\overline{\mathcal{B}}+\overline{U}(\mathcal{A}+i\mathcal{B})\right]\,.
\label{Vtilde}
\end{align}
\end{subequations}
As shown in
Appendix \ref{appA},  both matrices $V$ and $\tilde{V}$ are non-singular whenever $U$ is unitary.
Hence, we can write                                                                                                                                                                           
\begin{align}\label{6}
\vec{F}-i\vec{f}=\tilde{\mathcal{U}}(\vec{F}+i\vec{f})\,,
\end{align}
where we have defined $\tilde{\mathcal{U}}=V^{-1}\tilde{V}$.
Now, Eq.~\eqref{5} with $V$ and $\tilde{V}$ defined by Eq.~\eqref{Vs} implies that the matrix $\tilde{\mathcal{U}}$ is unitary
if and only if $U$ is a unitary matrix. This can be shown as follows. 
We note first  
that the unitarity of $V^{-1}\tilde{V}$ is equivalent to $VV^{\dagger}=\tilde{V}\tilde{V}^{\dagger}$ and find
\begin{align}\label{unitV}
VV^{\dagger}-\tilde{V}\tilde{V}^{\dagger}
= &2i\left[(\overline{\mathcal{A}}\mathcal{B}-\mathcal{A}\overline{\mathcal{B}})-\overline{U}(\overline{\mathcal{A}}\mathcal{B}-\mathcal{A}\overline{\mathcal{B}})\overline{U}^{\dagger}\right]\,.
\end{align}
We can show that $\overline{\mathcal{A}}\mathcal{B}-\mathcal{A}\overline{\mathcal{B}}=-i\mathbb{I}$ by
noting
\begin{align}\label{ginnerp}
g_\pm(a)\overline{g_\pm'(a)}-g_\pm'(a)\overline{g_{\pm}(a)} & = 
\frac{1}{2}
\int_{-a}^a \frac{d\ }{dx}\left[ g_{\pm}(x) \overline{g_\pm'(x)} - g'_\pm(x)\overline{g_\pm(x)}\right]dx\nonumber \\
&=\frac{1}{2}\left(\inpr{g_\pm}{A_U g_\pm}-\inpr{A_U g_\pm}{g_\pm}\right)\,,\nonumber \\
& = i\,.
\end{align}
By substituting this formula into Eq.~\eqref{unitV} we find
\begin{align}
VV^{\dagger}-\tilde{V}\tilde{V}^{\dagger} = 2(\mathbb{I}-\overline{U}\overline{U}^{\dagger})\,.
\end{align}
This equation shows that $\tilde{\mathcal{U}}$ is unitary, i.e., $VV^{\dagger}=\tilde{V}\tilde{V}^{\dagger}$,
if and only if $U$ is unitary.
It can also be shown that the map $U \mapsto \tilde{\mathcal{U}}$ is a bijection. 
The proof of this statement can be found in Appendix~\ref{appB}.  
Recall that the self-adjoint extensions $A_U$ are
uniquely 
parametrized by the unitary matrix $U$. We have seen that this unitary matrix is
in one-to-one correspondence with the unitary matrix $\tilde{\mathcal{U}}$.
Hence the self-adjoint extensions $A_U$ are in one-to-one correspondence with the set of boundary conditions~\eqref{6}
parametrized by the unitary matrix $\tilde{\mathcal{U}}$.

Note that Eq.~\eqref{6} is equivalent to
\begin{align}\label{revised-BC}
\begin{pmatrix} \tilde{f}'(a) - i \tilde{f}(a) \\ \tilde{f}'(-a) + i\tilde{f}(-a)\end{pmatrix} 
= \mathcal{U} \begin{pmatrix} \tilde{f}'(a) + i \tilde{f}(a) \\ \tilde{f}'(-a) - i\tilde{f}(-a)\end{pmatrix},
\end{align}
where the unitary matrix $\mathcal{U}$ is defined by 
\begin{align}
\mathcal{U} : = \frac{1}{2} \begin{pmatrix} 1 & 1 \\ -1 & 1\end{pmatrix} \tilde{\mathcal{U}}
\begin{pmatrix} 1 & -1 \\ 1 & 1\end{pmatrix}\,.
\end{align}
This set of boundary conditions is the same as those inferred using another methods for free quantum particle in 
a box~\cite{bonn}.  

In summary, in this section 
we have found the realization of the self-adjoint extensions of the differential operator~\eqref{OpA}
given by the domain extension~\eqref{SAdom} according to von~Neumann's theorem in terms of boundary conditions.
These boundary conditions are given by Eq.~\eqref{revised-BC}, where the $2\times 2$ unitary matrix $\mathcal{U}$ is
given by
\begin{align}
\mathcal{U} = - \frac{1}{2}\begin{pmatrix} 1 & 1 \\ -1 & 1\end{pmatrix}
[ \overline{\mathcal{A}} - i \overline{\mathcal{B}} + \overline{U}
(\mathcal{A} - i \mathcal{B})]^{-1}[\overline{\mathcal{A}}+i\overline{\mathcal{B}} + \overline{U}
(\mathcal{A}+i\mathcal{B})]\begin{pmatrix} 1 & - 1 \\ 1 & 1\end{pmatrix},
\end{align}
with the matrices $\mathcal{A}$ and $\mathcal{B}$ defined by Eq.~\eqref{ABM}.  We also have shown that the map
$U\mapsto \mathcal{U}$, where $U$ is the $2\times 2$ unitary matrix parametrizing the self-adjoint extensions given
by von~Neumann's theorem, is a bijection.  In the next section we rewrite the boundary conditions~\eqref{revised-BC} 
in a more familiar form, with those most frequently used as special cases.

\section{Classification of the boundary conditions}
\label{BoundaryConditions}
We first write the boundary conditions~\eqref{revised-BC} as follows:
\begin{align}\label{changed-BC}
(\mathbb{I} - \mathcal{U})\begin{pmatrix} \tilde{f}'(a) \\ \tilde{f}'(-a)\end{pmatrix}
= i (\mathbb{I} + \mathcal{U})\begin{pmatrix} \tilde{f}(a) \\ - \tilde{f}(-a)\end{pmatrix}\,.
\end{align}
To rewrite these boundary conditions in a more familar form, it is useful to classify them according to
whether or not the matrices $\mathbb{I}-\mathcal{U}$ or $\mathbb{I}+\mathcal{U}$ are singular. 
In this section we write $f(\pm a)$ instead of $\tilde{f}(\pm a)$ for simplicity.

\medskip

\noindent
\small{{\bf Case I: both $\mathbb{I}-\mathcal{U}$ and $\mathbb{I}+\mathcal{U}$ are regular}}.
In this case we can write Eq.~\eqref{changed-BC} as
\begin{align}\label{bc2}
\begin{pmatrix}
f'(a)\\[.5em]
f'(-a)
\end{pmatrix}&=
H
\begin{pmatrix}
f(a)\\[.5em]
-f(-a)
\end{pmatrix}\,,
\end{align}
where
\begin{align}\label{H}
H:=i(\mathbb{I}-\mathcal{U})^{-1}(\mathbb{I}+\mathcal{U})\,.
\end{align}
One can readily show that the $2\times 2$ matrix $H$ is Hermitian and invertible.  
Furthermore, the matrices $H$ and $\mathcal{U}$ commute and 
$\mathcal{U} = (H - i\mathbb{I})^{-1}(H+i\mathbb{I})$.  Thus, the invertible $2\times 2$ Hermitian matrix $H$ and the
$2\times 2$ unitary matrix $\mathcal{U}$, such that $\mathbb{I}\pm \mathcal{U}$ are non-singular, are in one-to-one
correspondence through the relation~\eqref{H}.

By writing
\begin{align}
H = \begin{pmatrix} \alpha & \beta \\ \overline{\beta} & - \gamma\end{pmatrix}\,,
\end{align}
where $\alpha,\gamma\in\mathbb{R}$, $\beta\in\mathbb{C}$ and $\alpha \gamma + |\beta|^2 \neq 0$,
we find that Eq.~\eqref{bc2} becomes
\begin{subequations}\label{bc3}
\begin{align}
f'(a)&=\alpha f(a)- \beta f(-a)\,,\label{bc3-1}\\
f'(-a)&=\overline{\beta} f(a)+\gamma f(-a)\,.\label{bc3-2}
\end{align}
\end{subequations}
Since the Hermitian matrix $H$ is invertible, one may also express the boundary conditions here by writing 
$f(a)$ and $f(-a)$ as linear combinations of $f'(a)$ and $f'(-a)$.
Notice that if $H$ is diagonal, 
which implies by virtue of Eq.~\eqref{H} that $\mathcal{U}$ is also diagonal, then Eqs.~\eqref{bc3-1}
and \eqref{bc3-2} become
$f'(a)=\alpha f(a)$ and
$f'(-a)=\gamma f(-a)$,
with $\alpha\neq 0$ and $\gamma\neq 0$.  These boundary conditions are called
the Robin boundary conditions.  

\medskip

\noindent
\small{{\bf  Case II. $\mathbb{I}+\mathcal{U}$ is singular and $\mathbb{I}-\mathcal{U}$ is regular.}}
This case is similar to case I and the boundary conditions are given by Eq.~\eqref{bc3} except that the Hermitian matrix $H$ is not invertible.  Thus, $f'(a)$ and $f'(-a)$ are proportional to each other as linear combinations of
$f(a)$ and $f(-a)$.
For the special case with  
$\mathcal{U}=-\mathbb{I}$ we have $H=0$, and  conditions \eqref{bc3} reduce to 
$f'(a)=f'(-a)=0$,
i.e.\ the Neumann boundary conditions at both $\pm a$.

\medskip

\noindent
\small{{\bf Case III. $\mathbb{I}-\mathcal{U}$ is singular and $\mathbb{I}+\mathcal{U}$ is regular.}}
In this case the matrix $H$ in Eq.~\eqref{bc2} is not defined. Instead we can write Eq.~\eqref{changed-BC} as
\begin{align}
\begin{pmatrix} f(a) \\ -f(-a)\end{pmatrix} = H'
\begin{pmatrix} f'(a) \\ f'(-a)\end{pmatrix}\,,
\end{align}
where
\begin{align}
H':=- i(\mathbb{I}+\mathcal{U})^{-1}(\mathbb{I}-\mathcal{U})\,,
\end{align}
which is Hermitian but not invertible.
By letting
\begin{align}
H' = \begin{pmatrix} \alpha' & - \beta' \\ -\overline{\beta'} & -\gamma'\end{pmatrix}\,,
\end{align}
we find
\begin{equations}
f(a)&=\alpha' f'(a)- \beta' f'(-a)\,,\\
f(-a)&=\overline{\beta}'f'(a)+\gamma' f'(-a)\,,
\end{equations}
with $\alpha',\gamma'\in\mathbb{R}$, and $\beta'\in\mathbb{C}$.
Since the matrix $H'$ is singular, $f(a)$ and $f(-a)$ are proportional to each other as linear combinations
of $f'(a)$ and $f'(-a)$.
For the special case with
$\mathcal{U}=\mathbb{I}$, that is, for $H'=0$, the boundary conditions reduce to
$f(a)=f(-a) = 0$,
namely, the Dirichlet boundary conditions at both $\pm a$.

\medskip

\noindent
\small{{\bf Case IV. Both $\mathbb{I}\pm \mathcal{U}$ are singular.}}
In this case
$\mathcal{U}$ has $1$ and $-1$ as eigenvalues. Then, $\mathcal{U}$ can be given as
\begin{align}
\mathcal{U} = \begin{pmatrix} \cos\theta & e^{-i\varphi}\sin\theta \\ e^{i\varphi}\sin\theta & - \cos\theta
\end{pmatrix}\,,
\end{align}
where $\theta \in [0,\pi]$ and $\varphi\in [0,2\pi)$ are the polar and azimuthal angles, respectively, in the
standard spherical polar coordinates.
(Thus, the matrix $\mathcal{U}$ is a Pauli spin matrix in the direction specified by the angles $\theta$ and
$\varphi$.)
By substituting this equation into Eq.~\eqref{changed-BC} we obtain
\begin{equations}
\sin\frac{\theta}{2}\left[f'(a)\sin\frac{\theta}{2} - f'(-a)e^{-i\varphi}\cos\frac{\theta}{2} \right]
& = i\cos\frac{\theta}{2}\left[f(a)\cos\frac{\theta}{2} - f(-a)e^{-i\varphi}\sin\frac{\theta}{2}\right]\,,\\
-\cos\frac{\theta}{2}\left[f'(a)e^{i\varphi}\sin\frac{\theta}{2} - f'(-a)\cos\frac{\theta}{2}\right] & = 
i\sin\frac{\theta}{2}\left[f(a)e^{i\varphi}\cos\frac{\theta}{2} - f(-a)\sin\frac{\theta}{2}\right]\,.
\end{equations}
These equations are equivalent to
\begin{equations}
f'(-a)\cos\frac{\theta}{2} & = f'(a)e^{i\varphi}\sin\frac{\theta}{2}\,, \label{BCBC1}\\
f(-a)\sin\frac{\theta}{2} & = f(a)e^{i\varphi}\cos\frac{\theta}{2}\,.\label{BCBC2}
\end{equations}
If $\theta \in (0,\pi)$, then we can write these boundary conditions as
\begin{equations}
f(-a) & = K f(a),,\\
f'(-a) & = \frac{1}{\overline{K}}f'(a)\,,
\end{equations}
where $K = e^{i\varphi}\cot\frac{\theta}{2}$ is any non-zero complex number.  For $\theta=\pi/2$ ($|K|=1$) we have $f(-a)=e^{i\varphi}f(a)$ and $f'(-a) = e^{i\varphi}f'(a)$. The boundary condition given by this set of equations
is often called an automorphic boundary condition.  In particular, if $\varphi=0$ ($K=1$), we have the periodic
boundary condition, whereas if $\varphi=\pi$, we have the anti-periodic boundary condition.

If $\theta=0$, then
Eqs.~\eqref{BCBC1} and \eqref{BCBC2} become
$f(a) = f'(-a) = 0$.
Thus, we have the Dirichlet boundary condition at $a$ and the Neumann boundary condition at $-a$. 
On the other hand, if $\theta=\pi$, then they become
$f(-a)= f'(a)=0$.
Thus, we have the Dirichlet boundary condiion at $-a$ and the Neumann  boundary condition at $a$.

\section{Conclusion}

In this paper we studied the self-adjoint extensions of the Schr\"odinger differential operator 
with a symmetric potential in one dimension with two boundaries.  These self-adjoint extensions according to
von~Neumann's theorem are parametrized by a $2\times 2$ unitary matrix.  On the other hand, the self-adjoint
extensions of this differential operator can be specified by the boundary conditions on the functions on which this
differential operator acts, also parametrized by a $2\times 2$ unitary matrix.  We expressed the latter matrix
explicitly in terms of the former and verified that they are in one-to-one correspondence.  This correspondence 
confirms that these boundary conditions give all possible self-adjoint extensions of this differential operator.

We also wrote down the boundary conditions parametrized by a $2\times 2$ matrix in a more familiar form.
In the classification given in Sec.~\ref{BoundaryConditions} 
we showed which particular choices of the unitary matrix in the parametrization of the self-adjoint extension correspond to familiar types of boundary conditions, such as Dirichlet, Neumann, Robin, periodic, anti-periodic and automorphic boundary conditions. 

As mentioned in Section \ref{Intro},
there are other systems with differential operators with self-adjoint extensions similar to the ones described in this work. Of particular interest is the case of the anti-de Sitter scalar field theory which involves a similar 
spatial differential operator, which we will investigate in a separate paper.

\begin{acknowledgements}

The authors would like to thank Chris Fewster for helpful discussions as well as Lasse Schmieding for numerous discussions and very useful insight he provided. This work was supported by 
Overseas Research Scholarship from the University of York.

\end{acknowledgements}

\appendix
\section{Non-singularity of the matrices $V$ and $\tilde{V}$}\label{appA}

In this appendix we prove that the matrices $V$ and $\tilde{V}$ defined in Eq.~\eqref{Vs} are invertible whenever the matrix $U$ is unitary, so that the matrix $\mathcal{U}=V^{-1}\tilde{V}$ exists.
Suppose that $V$ is singular.
Then there is a non-zero vector $\vec{a}=(a_1,a_2)^{T}\in\mathbb{C}^{2}$ such that $V\vec{a}=0$. Then, using the definition of $V$ in Eq.~\eqref{V}, we have
\begin{align}
\|(\overline{\mathcal{A}}-i\overline{\mathcal{B}})\,\vec{a}\|^{2}=\|(\mathcal{A}-i\mathcal{B})\,\vec{a}\|^{2}\,,
\end{align}
where we have used the fact that $U$ is unitary. 
By substituting the expressions for the matrices $\mathcal{A}$ and $\mathcal{B}$ in Eq.~\eqref{ABM}, we find
\begin{align}
&|\overline{g_{+}(a)}-i\overline{g_{+}'(a)}|^{2}|a_{1}|^{2}
+|\overline{g_{-}(a)}-i\overline{g_{-}'(a)}|^{2}|a_{2}|^{2}\nonumber\\
&=|g_{+}(a)-ig_{+}'(a)|^{2}||a_{1}|^{2}+|g_{-}(a)-ig_{-}'(a)|^{2}|a_{2}|^{2}\,.\label{mods}	
\end{align}
Then by using Eq.~\eqref{ginnerp} we find $|a_1|^{2}+|a_2|^{2}=0$, which contradicts the assumption that
$\vec{a}\neq \vec{0}$.  Hence, $V$ cannot be a singular matrix.  The proof is similar for $\tilde{V}$.

\section{Proof that the map $U\mapsto\mathcal{U}=V^{-1}\tilde{V}$ is a bijection}\label{appB}
We have shown Sec.~\ref{Body} that if the matrix $U$ is unitary, then there is a unique unitary matrix $\mathcal{U}$
satisfying $V=\mathcal{U}\tilde{V}$, where the matrices $V$ and $\widetilde{V}$ are defined by Eqs.~\eqref{V} and \eqref{Vtilde}, respectively. We now show the 
converse: if $\mathcal{U}$ is any unitary matrix, there is a unique unitary matrix $U$ such that 
$V=\mathcal{U}\tilde{V}$ . Note that if there is a matrix $U$ such that this equation is satisfied, then it must be unitary by the result in Sec.~\ref{Body}.

We first note that the relation  $V = \mathcal{U}\tilde{V}$ may be written as
\begin{align}\label{to-solve}
\overline{U}(\mathcal{A} - i\mathcal{B}) - \mathcal{U}\overline{U} (\mathcal{A}+i\mathcal{B}) = - \overline{A}-i\overline{B} + \mathcal{U}(\overline{A}+i\overline{B})\,.
\end{align}
This matrix equation can be regarded as simultaneous linear equations in the elements of 
the matrix $\overline{U}:=(u_{ij})$, with $1\leq i,j\leq 2$. Hence, 
Eq.~\eqref{to-solve} admits a unique solution if and only if the equation
\begin{align}\label{zero-eq}
\overline{U}(\mathcal{A} - i\mathcal{B}) - \mathcal{U}\overline{U} (\mathcal{A}+i\mathcal{B}) = 0\,,
\end{align}
implies $\overline{U}=0$. 
We now show that this is indeed the case.
 By moving the second term in Eq.~\eqref{zero-eq}
to the right-hand side and multiplying by the adjoint from the left, we find
\begin{align}
(\overline{\mathcal{A}}+i\overline{\mathcal{B}})\overline{U}^\dagger \overline{U} (\mathcal{A}-i\mathcal{B})
= (\overline{\mathcal{A}} - i\overline{\mathcal{B}}) \overline{U}^\dagger \overline{U} (\mathcal{A}+i\mathcal{B}),
\end{align}
i.e.,
\begin{align}
i(\overline{\mathcal{B}}\overline{U}^\dagger \overline{U} \mathcal{A} - 
\overline{\mathcal{A}}\overline{U}^\dagger \overline{U}\mathcal{B}) = 0\,.
\end{align}
By using the definitions of $\mathcal{A}$ and $\mathcal{B}$ in Eq.~\eqref{ABM} and the relation~\eqref{ginnerp},
we find that the diagonal elements of this matrix equation are
$|u_{11}|^2 + |u_{21}|^2 = 0$ and $|u_{12}|^2 + |u_{22}|^2 = 0$, which imply
$U=0$.

\section{The case with a non-symetric real-valued potential}\label{appC}
In this appendix we show that the self-adjoint extensions parametrized by
a $2\times 2$ unitary matrix given by von~Neumann's theorem are in one-to-one correspondence with
the boundary conditions~\eqref{revised-BC} even for a non-symmetric potential, with which the eigenfunctions
of the operator $A$ given by Eq.~\eqref{OpA} cannot be chosen to have definite parity.  In this case, however,
the relation between the unitary matrix $U$ parametrizing the self-adjoint extensions given by von~Neumann's theorem 
and those characterizing the boundary conditions, $\mathcal{U}$, is rather implicit as we shall see.

Let $g_1(x)$ and $g_2(x)$ be two orthonormal solutions, i.e., solutions satisfying
$\langle g_i,g_j\rangle = \delta_{ij}$, of the equation,
\begin{align}\label{A2}
A_Ug_j(x) = i g_j(x)\,,
\end{align}
with $A_U$ being again a self-adjoint extension of the operator $A$ given 
by Eq.~\eqref{OpA} without the condition that $V(x)$ be symmetric.
We can then show, using the equation $\langle g_j,A_U g_k\rangle - \langle A_U g_j,g_k\rangle = 2i\delta_{jk}$, that
\begin{align}\label{othg}
\overline{g_j'(a)}g_k(a) - \overline{g_j(a)}g_k'(a)
- \overline{g_j'(-a)}g_k(-a) + \overline{g_j(-a)}g_k'(a) & = 2i\delta_{jk}\,.
\end{align}
We also note that since $\langle \overline{g_j}, Ag_k\rangle - \langle A \overline{g_j},g_k\rangle = 0$, we have
\begin{align}\label{othg2}
g_j'(a)g_k(a) - g_j(a)g_k'(a)
- g_j'(-a)g_k(-a) + g_j(-a)g_k'(a)  = 0\,,
\end{align}
which is trivially satisfied if $j=k$.

Defining the functions $G_{j}(x)$, $j=1,2$, in the same way as in Eq.~\eqref{G}, we may
again impose the symmetry condition on the functions in the domain of $A_{U}^{\dagger}$, which leads to the same
expression as Eq.~\eqref{G1}. 
This equation can be rearranged as
\begin{align}\label{mmm}
&
\left[ \overline{G_j'(a)}+i\overline{G_j(a)}\right]\left[ f'(a) + i f(a)\right] + \left[ \overline{G_j(-a)} - i\overline{G_j(-a)}\right]
\left[ f'(-a) - i f(-a)\right] \nonumber \\
& = \left[ \overline{G_j'(a)}-i\overline{G_j(a)}\right]\left[ f'(a) - i f(a)\right] + 
\left[ \overline{G_j(-a)} + i\overline{G_j(-a)}\right]
\left[ f'(-a) + i f(-a)\right]. 
\end{align} 
It will be convenient to define the following vectors:
\begin{align}\label{z-definition}
\vec{z}_j^{\,\,(+)} :=  \begin{pmatrix} G_j'(a) - i G_j(a) \\ G_j'(-a) + i G_j(-a) \end{pmatrix}, \hspace{.5cm}
\vec{z}_j^{\,\,(-)} :=  \begin{pmatrix} G_j'(a) + i G_j(a) \\ G_j'(-a) - i G_j(-a)\end{pmatrix}\,,
\end{align}
and 
\begin{align}
\vec{F}^{\,(+)}  :=  \begin{pmatrix} f'(a) + if(a) \\ f'(-a) - i f(-a) \end{pmatrix} \hspace{.5cm}
\vec{F}^{\,(-)}  : = \begin{pmatrix} f'(a)-if(a) \\ f'(-a) + i f(-a)\end{pmatrix}.
\end{align}
Then Eq.~\eqref{mmm} can be written as
\begin{align}\label{zF-relation}
( \vec{z}_j^{\,\,(+)}, \vec{F}^{\,(+)})_{\mathbb{C}^2} = (\vec{z}_j^{\,\,(-)}, \vec{F}^{\,(-)})_{\mathbb{C}_2},
\end{align}
where $(\cdot,\cdot)_{\mathbb{C}^2}$ is the standard inner product on $\mathbb{C}^2$.

First we show that each of the sets $\{\vec{z}^{\,\,(+)}_1, \vec{z}^{\,\,(+)}_2\}$ and
$\{\vec{z}^{\,\,(-)}_1, \vec{z}^{\,\,(-)}_2\}$ is linearly independent 
if the matrix $U$ is unitary.
Using the definition of $G_j(x)$, Eq.~\eqref{G} 
and the relations~\eqref{othg} and \eqref{othg2}
and their complex conjugates, it can be shown that
\begin{align}\label{zj-zk}
( \vec{z}_j^{\,\,(\pm)},\vec{z}_k^{(\,\,\pm)})_{\mathbb{C}^2}
 =&  \overline{G'_j(a)}G'_k(a) + \overline{G_j(a)}G_k(a)
+ \overline{G'_j(-a)}G'_k(-a) + \overline{G_j(-a)}G_k(-a) \nonumber\\
&\pm (\delta_{kj}- (UU^\dagger)_{kj})\,,
\end{align}
where $(UU^\dagger)_{kj}$ is the $kj$-element of the matrix $UU^\dagger$.
Then, the unitarity of the matrix $U$ implies
\begin{align}
( \vec{z}_j^{\,\,(\pm)},\vec{z}_k^{(\,\,\pm)})_{\mathbb{C}^2}
& =  \overline{G'_j(a)}G'_k(a) + \overline{G_j(a)}G_k(a)
+ \overline{G'_j(-a)}G'_k(-a) + \overline{G_j(-a)}G_k(-a)\,.
\end{align}
This implies that the linear transformation defined by
$\vec{z}_j^{\,\,(\sigma)} \mapsto \begin{pmatrix} G_j'(a) & G_j(a) & G_j'(-a) & G_j(-a)\end{pmatrix}^T$, $j=1,2$,
for $\sigma=+$ or $-$, is an isometry.  Hence, $\vec{z}_1^{\,\,(\sigma)}$ and $\vec{z}_2^{\,\,(\sigma)}$ are
linearly dependent if and only if $\begin{pmatrix} G_j'(a) & G_j(a) & G_j'(-a) & G_j(-a)\end{pmatrix}^T$, $j=1,2$, are.
 However, this would be impossible because from Eqs.~\eqref{othg} and
\eqref{othg2} it follows that
\begin{align}
\overline{g'_j(a)}G_k(a) - \overline{g_j(a)}G'_k(a)
- \overline{g'_j(-a)}G_k(-a) + \overline{g_j(-a)}G_k'(-a)
 =  2i\delta_{jk}\,,
\end{align}
which cannot be satisfied if the vectors
$\begin{pmatrix} G_j'(a) & G_j(a) & G_j'(-a) & G_j(-a)\end{pmatrix}^T$, $j=1,2$, were linearly dependent.
Thus, each of the sets $\{ \vec{z}^{\,\,(+)}_1, \vec{z}^{\,\,(+)}_2\}$ and
$\{ \vec{z}^{\,\,(-)}_1, \vec{z}^{\,\,(-)}_2\}$ is linearly independent.

Now, Eq.~\eqref{zj-zk} implies that 
$(\vec{z}_j^{\,\,(+)}, \vec{z}_k^{\,\,(+)})_{\mathcal{C}^2} = (\vec{z}_j^{\,\,(-)},\vec{z}_k^{\,\,(-)})_{\mathbb{C}^2}$
if and only if $U$ is unitary.  
In other words $\vec{z}_j^{\,\,(-)} = \mathcal{U}^\dagger \vec{z}_j^{\,\,(+)}$, where $\mathcal{U}$ is unitary, if and only if $U$ is unitary.

The unitary matrix $\mathcal{U}$ is uniquely determined for each $U$ because the vectors $\vec{z}_j^{\,\,(\pm)}$ satisfying $(\vec{z}_j^{\,\,(+)},\vec{z}_k^{\,\,(+)})_{\mathbb{C}^2} = (\vec{z}_j^{\,\,(-)},\vec{z}_k^{\,\,(-)})_{\mathbb{C}^2}$ uniquely
determine $\mathcal{U}$. We show next that, conversely, the unitary matrix $\mathcal{U}$ 
uniquely determines the matrix $U$ 
through the equation $\vec{z}_j^{\,\,(-)} = \mathcal{U}^\dagger\vec{z}_j^{\,\,(+)}$.
By substituting the defintions~\eqref{z-definition} of $\vec{z}_j^{\,\,(\pm)}$ and then the definitions~\eqref{G}
of $G_j(x)$, and writing the $jk$-element of $U$ as $u_{jk}$,  we find
\begin{align}
 &\begin{pmatrix} u_{j1}(\overline{g'_1(a)}+ i\overline{g_1(a)})  + u_{j2}(\overline{g_2'(a)}
+i\overline{g_2(a)})\\
u_{j1}(\overline{g'_1(-a)} - i \overline{g_2(-a)}) + u_{j2}(\overline{g'_2(-a)} - i
\overline{g_2(-a)})\end{pmatrix}\nonumber\\
& = \mathcal{U}
  \begin{pmatrix} u_{j1}(\overline{g'_1(a)}- i\overline{g_1(a)})  + u_{j2}(\overline{g_2'(a)}-i\overline{g_2(a)}) \\
u_{j1}(\overline{g'_1(-a)} + i \overline{g_2(-a)}) + u_{j2}(\overline{g'_2(-a)} + i\overline{g_2(-a)})\end{pmatrix} + \vec{v}\,,
\end{align}
where $\vec{v}$ is a $2$-dimensional column vector independent of $u_{jk}$. 
What we need to show is that the homogeneous equation obtained by
setting $\vec{v}=\vec{0}$ has the unique solution $u_{j1}=u_{j2}=0$ for $j=1,2$.  
Since the matrix $\mathcal{U}$ is unitary, the homogeneous equation is satisfied only if
\begin{align}
& |\overline{u_{j1}}(g'_1(a)- ig_1(a))  + \overline{u_{j2}}(g_2'(a)-ig_2(a))|^2  + |\overline{u_{j1}}(g'_1(-a) + i g_2(-a)) + \overline{u_{j2}}(g'_2(-a) + ig_2(-a))|^2 \nonumber \\
&= |u_{j1}(g'_1(a)- ig_1(a))  + u_{j2}(g_2'(a)-ig_2(a))|^2 + |u_{j1}(g'_1(-a) + i g_2(-a)) + u_{j2}(g'_2(-a) + ig_2(-a))|^2\,.
\end{align}
This equation can be simplified as
\begin{align}
\sum_{k,m}\overline{u_{jk}}u_{jm}(\overline{g_{k}'(a)}g_{m}(a)-\overline{g_{k}(a)}g_{m}'(a)-\overline{g_{k}'(-a)}g_{m}(-a)+\overline{g_{k}(-a)}g_{m}'(-a)) = 0\,.
\end{align}
By Eq.~\eqref{othg}
this equation reduces to $|u_{j1}|^2 + |u_{j2}|^2 = 0$.  That is, $u_{11}=u_{12}=u_{21}=u_{22}$.  Therefore,
the unitary matrix $\mathcal{U}$ uniquely determines $U$ (and vice versa).

Now, by substituting the equation $\vec{z}^{\,\,(-)}_j = \mathcal{U}^\dagger\vec{z}^{\,\,(+)}$ into 
Eq.~\eqref{zF-relation}, we have
\begin{align}
(\vec{z}^{\,\,(+)}_j, \vec{F}^{\,\,(+)})_{\mathbb{C}^2} 
= (\vec{z}_j^{\,\,(+)},\mathcal{U}\vec{F}^{\,\,(-)})_{\mathbb{C}^2}\,.
\end{align}
Since the inner product with two linearly independent vectors uniquely determines a vector, we have
$\vec{F}^{\,\,(+)} = \mathcal{U}\vec{F}^{\,\,(-)}$, i.e.
\begin{align}\label{bcnonsym}
\begin{pmatrix} f'(a) + if(a) \\ f'(-a) - i f(-a) \end{pmatrix} 
= \mathcal{U}\begin{pmatrix} f'(a)-if(a) \\ f'(-a) + i f(-a)\end{pmatrix}\,.
\end{align}
Thus, we have shown that the self-adjoint extensions parametrized by a $2\times 2$ unitary matrix are in one-to-one
correspondence with the boundary conditions given by this equation parametrized by another $2\times 2$ unitary
matrix $\mathcal{U}$.

\bibliography{references}

\end{document}